# Multiple-level Green Noise Mask Design for Practical Fourier Phase Retrieval

Qiuliang Ye*, *Student Member, IEEE,* Bingo Wing-Kuen Ling, *Senior Member, IEEE,* Li-Wen Wang, *Member, IEEE,* and Daniel Pak-Kong Lun*, *Senior Member, IEEE*

*Abstract*—Phase retrieval, a long-established challenge for recovering a complex-valued signal from its Fourier intensity measurements, has attracted significant interest because of its far-flung applications in optical imaging. To enhance accuracy, researchers introduce extra constraints to the measuring procedure by including a random aperture mask in the optical path that randomly modulates the light projected on the target object and gives the coded diffraction patterns (CDP). It is known that random masks are non-bandlimited and can lead to considerable high-frequency components in the Fourier intensity measurements. These high-frequency components can be beyond the Nyquist frequency of the optical system and are thus ignored by the phase retrieval optimization algorithms, resulting in degraded reconstruction performances. Recently, our team developed a binary green noise masking scheme that can significantly reduce the high-frequency components in the measurement. However, the scheme cannot be extended to generate multiple-level aperture masks. This paper proposes a two-stage optimization algorithm to generate multi-level random masks named *OptMask* that can also significantly reduce high-frequency components in the measurements but achieve higher accuracy than the binary masking scheme. Extensive experiments on a practical optical platform were conducted. The results demonstrate the superiority and practicality of the proposed *OptMask* over the existing masking schemes for CDP phase retrieval.

## I. Introduction

PHASE retrieval aims to reconstruct a complex-valued optical wavefield from intensity-only measurements. It is known that, in coherent optical systems, the phase changes of the receiving light carry a lot of information about the measuring object that cannot be found in the intensity of the light. However, traditional imaging devices (charge-coupled device (CCD) or complementary metal-oxide-semiconductor (CMOS) sensors) can only detect the intensity of optical waves but not the phase information [1]. The need for phase retrieval algorithms thus naturally arises. Phase retrieval is a key problem in crystallography, optical imaging, astronomy, X-ray, electronic imaging [2]–[8], etc. Recently, substantial progress was made in the development of phase retrieval algorithms due to the advance in optimization theories [1]. Specifically, optical masks (also named coded apertures), acting as an extra constraint to the optimization process, are adopted in optical phase retrieval systems to improve reconstruction performance. It is shown in [9], [10] that the introduction of the random masks significantly improves the accuracy of the reconstructed signals.

The most representative approach in this category is called the coded diffraction pattern (CDP) [9], [10] scheme which aims to reconstruct a complex-valued signal $\mathbf{x} \in \mathbb{C}^n$ from its modulated Fourier intensity measurements $\mathcal{X}_l = |\mathcal{F}(\mathbf{T}_l \circ \mathbf{x})|^2$, $l = 1, \ldots, L$, where $L$, $\mathbf{T}$ and $\circ$ denote the number of measurements, optical masks, and elementwise multiplication, respectively. $\mathcal{F}$ represents the Fourier transform operator. In practice, the Fourier transform is implemented by optical devices. The optical masks can be realized through optoelectronic devices like spatial light modulators (SLM) or digital micromirror devices (DMD) [11]–[14], or by fabrication [15].

An example of the SLM-based CDP phase retrieval system (optical path) is shown in Fig. 1. The SLM, which realizes the random mask $\mathbf{T}$, is used to modulate the phase or amplitude of the incident light beam that is projected onto the target object $\mathbf{x}$. The light then passes through a lens that performs the Fourier transform $\mathcal{F}$ optically and is captured by the imaging sensor to become the Fourier intensity measurement $\mathcal{X}$ of the object. For CDP phase retrieval, the significance of coded apertures' randomness is highlighted in many research articles to guarantee reconstruction performance [9], [16], [17]. Multiple-level random masks are widely employed to ensure randomness.

Manuscript received ?? ??, 2023; revised ??? ??, 2022 and ??? ??, 2022; accepted ??? ??, 2022. Date of publication ??? ??, 2022; date of current version ??? ??, 2023. The work presented in this article was supported by the Hong Kong Research Grant Council under General Research Fund no. PolyU 15225321. The associate editor coordinating the review of this manuscript and approving it for publication was ????? ????? ?????. (Corresponding author: Qiuliang Ye and Daniel Pak-Kong Lun.)

Qiuliang Ye, Li-Wen Wang, and Daniel Pak-Kong Lun are with the Department of Electronic and Information Engineering, The Hong Kong Polytechnic University, Kowloon, Hong Kong SAR, China. Bingo Wing-Kuen Ling is with the School of Information Engineering, Guangdong University of Technology, Guangdong Province, China. (E-mail: qiuliang.ye@connect.polyu.hk and pak.kong.lun@polyu.edu.hk)

Digital Object Identifier

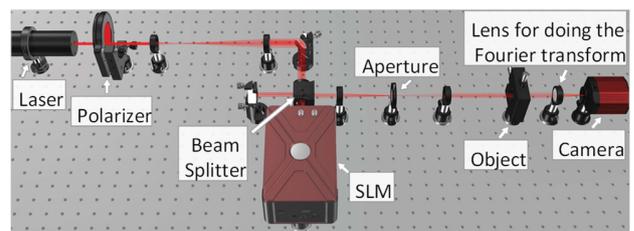

Fig. 1: The optical path of the SLM-based CDP phase retrieval system.

However, random masks are non-bandlimited in principle [18]. The resulting modulated Fourier intensity measurements contain a large number of significant high-frequency components. Fig. 2 shows an example of Fourier intensity measurements of an object when a white noise mask [9] is used. The image on the left was generated by computer simulation and the image on the right was collected from a practical CDP-based phase retrieval system. The red square shows the boundary of the 0-th diffraction order, which is also the Nyquist frequency of the masked signal $\mathbf{T}_l \circ \mathbf{x}$. It can be seen that there are many data of significant values outside the red square. Note that all phase retrieval optimization methods are discrete algorithms. They assume that the masked signal $\mathbf{T}_l \circ \mathbf{x}$ is discrete with the sampling rate defined by the pixel pitch of the SLM. Since it is discrete, its spectrum will have a Nyquist frequency, which is the red square in Fig. 2. All discrete systems can only handle the data up to the Nyquist frequency. Thus, the data outside the red square in Fig. 2 will be ignored. It is equivalent to lowpass filtering the Fourier intensity measurements, and it would bring errors and thus inevitably degrade the reconstruction performance. We will demonstrate the effect in Section VI.

In recent years, various masking schemes have been proposed for CDP-based phase retrieval systems [9], [16], [17]. However, they only focused on recovery guarantees and verified the approaches in simulated environments. In simulation, the Fourier transform is always approximated by the discrete Fourier transform (DFT). The resulting Fourier intensity measurements thus will not have any data outside the Nyquist frequency even when non-bandlimited masks are used. The problem is thus unnoticed. Recently, our team revealed the problem through optical experiments and proposed a green noise masking scheme to mitigate the problem [18]. By using a half-toning technique [19], we designed binary masks with energy concentrated in the mid-frequency band and applied them to CDP phase retrieval systems. Note that the use of bandlimited masks does not affect the frequency response of the reconstructed images but only avoids the generation of extremely high-frequency data that are beyond the Nyquist frequency of the system. Thus, the performance of the reconstruction is improved. However, since the masks have only binary values, there can be many sharp changes between adjacent pixels which inevitably introduce high-frequency components to the mask. In fact, these sharp changes can be smoothed out if the pixel values of the mask have multiple levels. In this paper, we propose a novel multi-level random phase mask named *OptMask* for CDP phase retrieval. Besides, we also propose a new optimization-based method for designing *OptMask*. Compared to the previous binary amplitude masking scheme, the proposed design method allows the flexibility for the user to design *OptMask* to have the energy concentrated in the low to mid-frequency band while maximizing its randomness. It is achieved by fully making use of the masks' multiple-level pixel values. *OptMask* is a phase mask. It is different from the binary amplitude masks that will block the light when the mask value is 0. Hence, *OptMask* allows the maximal amount of light to project on the object. *OptMask* can be implemented with commercial SLM devices or by fabrication, hence they are readily available for practical applications. To summarize, the contributions of this work are as follows:

1. We propose a novel multiple-level green noise phase mask named *OptMask* for CDP phase retrieval. The mask fully makes use of its multiple-level and bandpass properties to increase its randomness and reduce the extremely high-frequency components in the Fourier intensity measurements that are beyond the Nyquist frequency of the optical system.
2. We propose a novel two-stage optimization algorithm for the design of *OptMask*. The method makes use of a two-stage optimization technique to let *OptMask* possess all required properties for efficient implementation in practical optical systems.
3. We verify *OptMask* with a practical CDP Fourier phase retrieval platform to demonstrate its practicality and show its effectiveness over the existing masks. A comprehensive analysis is also provided for verifying the optimality of *OptMask*.

The rest of this paper is organized as follows: Section II reviews the CDP framework and related mask design methods. We propose a two-stage optimization algorithm for designing *OptMask* in Section III and provide a comprehensive analysis of the parameters for *OptMask* in Section IV. In Section V, we provide extensive experimental results on an optical system for validating the performance of the proposed masking scheme and compare it with the existing approaches. The conclusion is drawn in Section VI.

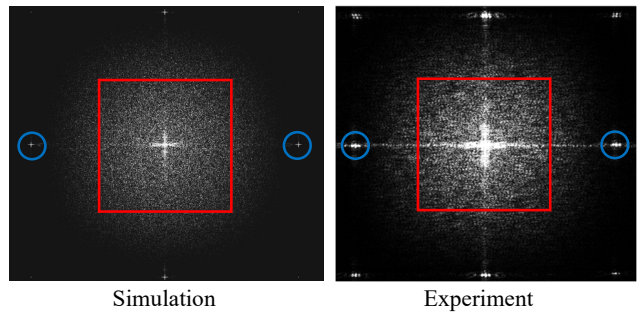

Simulation    Experiment

Fig. 2: Fourier intensity measurements (left: simulation; right: experiment) of a USAF chart multiplied with the same white noise (normal distribution) mask. The contrast of the images is adjusted to visualize the small coefficients. The red square denotes the 0-th diffraction order. The blue circles are the central regions of the 1-st diffraction order.

## II. REVIEW OF CDP FRAMEWORK AND MASK DESIGN METHODS

### A. Coded Diffraction Pattern

The CDP phase retrieval framework recovers an unknown complex-valued signal $\mathbf{x}$ from its randomly modulated Fourier intensities [9]. Since it is mainly studied in a discrete configuration, the Fourier intensities are approximated by the intensities of DFT. To be specific, in the 2D phase retrieval problem, which is the main interest of this paper, the value at $[k_1, k_2]$ of the $l$-th intensity measurement $\mathcal{X}_l$ is given by:

$$\left| \frac{1}{\sqrt{M_1 M_2}} \sum_{n_1,n_2=0}^{N_1,N_2} x[n_1,n_2] T_l[n_1,n_2] e^{-i2\pi\left(\frac{k_1 n_1}{M_1} + \frac{k_2 n_2}{M_2}\right)} \right|^2 \quad (1)$$

where $M_i \geq N_i$ for $i = 1,2$ and $0 \leq k_i \leq M_i - 1$. $\mathbf{T}_l$ denotes the masks (coded apertures) and $l = 1, \ldots, L$, where $L$ refers to the number of masks. $\mathcal{X}$ is called the *coded diffraction pattern* since it provides information about the spectrum of $\mathbf{x}$ modulated by the mask $\mathbf{T}$. The CDP framework provides the uniqueness guarantee (up to a unimodular constant) with a sufficient number of measurements (empirically larger than 4) and *random* masks that obey the *"admissible"* property [9].

In practice, $\mathbf{T}$ can be realized through diffractive optical elements like spatial light modulators (SLMs) or digital micromirror devices (DMDs), which contain millions of independent cells (pixels); each of them can modulate the amplitude or phase of the light projected to it. Due to the nature of the device, DMD can only realize the binary amplitude modulation (0 and 1) while SLM can create multiple-level phase shifts (usually 256 levels). An example of the CDP phase retrieval system is shown in Fig. 1.

*B. Non-bandlimited Random Mask*

The CDP phase retrieval deals with the Fourier intensity measurements which are also named Fraunhofer diffraction patterns in optics. According to the optical diffraction theory [20], the complex-valued wave field $\mathbf{O}$ through wavefront propagation from an object $\mathbf{o}$ at the Fourier plane is given by the Fraunhofer transform:

$$O(k_u, k_v) \approx C_f^\lambda \iint o(u,v) e^{j2\pi\left[\frac{p(uk_u + vk_v)}{\lambda f}\right]} du dv, \quad (2)$$

where $f$ is usually denoted as the focal length of the lens, and $p$ is the pixel pitch of the SLM or DMD. $(u,v)$ and $(k_u, k_v)$ represent the spatial and spectral coordinates, respectively, and $C_f^\lambda$ is a constant related to the wavelength $\lambda$ and focal length $f$. Considering the object $\mathbf{o} = \mathbf{x} \circ \mathbf{T}_l$, $l = 1, \ldots, L$ in this research, (2) can be reformulated as:

$$O(k_u, k_v) \approx \bar{C}_f^\lambda \mathcal{F}\left\{ x\left(\frac{pu}{\lambda f}, \frac{pv}{\lambda f}\right) \right\} * \mathcal{F}\left\{ T\left(\frac{pu}{\lambda f}, \frac{pv}{\lambda f}\right) \right\} \quad (3)$$

where $*$ denotes the convolution integral, and $\bar{C}$ is a constant. As can be seen, the wave field $\mathbf{O}$ is the convolution between the Fourier transform of $\mathbf{x}$ and $\mathbf{T}$ with the scaling factor $\frac{\lambda f}{p}$. The central $\left(-\frac{\lambda f}{2p}, \frac{\lambda f}{2p}\right)$ of the Fourier plane is called the 0-th diffraction order, which is indicated by the red square in Fig. 2. The pattern will repeat itself (but modulated with an envelope) to form the higher-order diffractions although with lower amplitude. Note that $\mathcal{F}\{\mathbf{T}\}$ is non-bandlimited if $\mathbf{T}$ is random. Therefore, the convolution of $\mathcal{F}\{\mathbf{x}\}$ and $\mathcal{F}\{\mathbf{T}\}$ is also non-bandlimited [21]. As is shown in Fig. 2, there are many high-frequency data outside the 0-th diffraction order.

The objective of the CDP phase retrieval algorithm is to reconstruct the target images from their corresponding intensities [9], [10]:

$$\text{Find } \mathbf{x} \in \mathbb{C}^{N_1 \times N_2} \quad (4)$$

s.t. $\mathcal{X}_l = |\mathcal{F}(\mathbf{T}_l \circ \mathbf{x})|^2$, $l = 1, \ldots, L$.

For efficient computation, all CDP phase retrieval algorithms assume $\mathbf{T}_l \circ \mathbf{x}$ are discrete samples and approximate the Fourier transform $\mathcal{F}$ by DFT. Based on the sampling theory, the Fourier intensity given by the DFT of discrete samples is periodic and has a Nyquist frequency defined by the sampling rate of the discrete data. In this case, it is defined by the sampling rate of $\mathbf{T}_l \circ \mathbf{x}$, which is determined by the pixel pitch of the mask $\mathbf{T}$. The Nyquist frequency thus coincides with the 0-th diffraction order boundary, as shown by the red square in Fig. 2. Since the CDP phase retrieval algorithm is a discrete algorithm, it can only handle the data up to the Nyquist frequency. Therefore, the high-frequency components of the intensity measurements that are beyond the red square in Fig. 2 will not be considered in the optimization algorithm. It is equivalent to filtering out these high-frequency components and thus leads to degraded reconstruction performances.

*C. Masking Schemes for CDP Phase Retrieval*

In recent years, different masking schemes for CDP phase retrieval were proposed to provide theoretical recovery guaranteed and improve the reconstruction performance under noisy situations. The masking schemes can be broadly categorized into amplitude masks, amplitude-phase masks, and phase masks [9], [16]–[18].

An amplitude mask modulates the amplitude of the incident beam through passing, blocking, or attenuating photons. Binary masks are commonly used because of their practicality. They can be implemented by DMD or fabrication. Several binary masking schemes were proposed for phase retrieval, e.g., white-noise masks [9], blue-noise masks [22], and green-noise masks [18] (all $\mathbf{T} \in \{0,1\}$). However, a significant amount of photons are lost when using the amplitude masks, thus leading to lower signal-to-noise-ratio (SNR) measurements [15]. It is undesirable for photon-limited imaging systems like fluorescence imaging. This greatly limits the applicability of the amplitude masks.

An amplitude-phase mask adjusts the amplitude and phase of the incident beam. However, it cannot modulate the amplitude and phase simultaneously. Therefore, two devices and two 4F (4 focal-length) systems are needed to achieve amplitude phase modulation. This inevitably increases the difficulty of the optical alignment and the cost of the whole system. Besides, the state-of-the-art amplitude-phase masking schemes allow the elements to give a magnitude larger than 1 [9], [23], which violates the energy preservation condition and further complicates the system implementation.

A phase mask changes the phase of the incident light. It can be implemented by the phase-only SLM. The phase mask provides high SNR since it allows the light to get through it without blocking. Phase masks were popularly adopted in the CDP phase retrieval problem. In [9], discrete uniformly distributed random variables $\mathbf{T} \in \{1, -1, -i, i\}$ were used to form the coded apertures. They have a flat frequency response (white noise). Taking temporal correlation and spatial separation into consideration, [16] developed the uniform random masks $\mathbf{T} \in \{1, -1, -i, i\}$ that satisfy the "admissible" condition. Aiming to minimize the upper bounds of the Gershgorin theorem, [17] developed the uniform random masks $\mathbf{T} \in \{1, -1, -i, i\}$ based on an unconstrained optimization algorithm. However, these masking schemes did not consider the non-bandlimited property of the random masks and their adverse effect on practical CDP phase retrieval systems, as described in Section II-B. Recently, our team proposed using the binary phase-only green-noise mask $\mathbf{T} \in \{1, -1\}$ with energy concentrated in the mid-frequency band to mitigate the problem due to non-bandlimited random masks [18]. However, the error diffusion algorithm for generating the green-noise masks was originally developed for half-toning. It was not optimally designed for the phase retrieval task. Besides, since it is a binary mask, there can be many sharp changes between adjacent pixels. Such sharp changes will introduce high-frequency components to the mask, which in turn degrades the phase retrieval performance. In this paper, we propose a multiple-level green noise phase mask called *OptMask* that can get around the above-mentioned problems. The details are discussed in the following sections.

### III. MULTI-LEVEL RANDOM MASKS AND TWO-STAGE OPTIMIZATION METHOD FOR MASK DESIGN

#### A. Design Considerations

In this section, the design considerations of the proposed *OptMask* are explained. From [9], it is known that the randomness of the mask is the key to the performance of phase retrieval when using the CDP method. The randomness ensures the statistical independence of each measurement thus leading to better reconstruction performance. Therefore, we expect that *OptMask* is sampled from a probability distribution. On the other hand, the non-bandlimited nature of random masks explained in Section II-B can lead to degradation in performance. *OptMask* should have as less high-frequency components as possible while maintaining randomness.

Common imaging devices have a precision limited to 12 to 16 bits. When they are used to record the Fourier intensity, we need to ensure the Fourier coefficients are within the dynamic range of the imaging devices. It is known in (3) that the Fourier intensity measurement is the convolution between the Fourier transforms of the object and the mask. While we cannot control the magnitude of the object's Fourier coefficients, we should minimize the magnitude of the mask's Fourier coefficients. In particular, we need to ensure that *OptMask* will not have a bias in the spatial domain or that its Fourier transform will have a strong peak around the DC term. It will saturate the imaging device and lead to distortion. Besides, we also need to control the number of quantization levels of *OptMask* so that it will not generate very fine data value variations that cannot be implemented or detected by the optical devices. To summarize, *OptMask* should be a phase mask with the following characteristics:
1. Randomness: the masks should be as random as possible.
2. Bandpass: the masks should have a bandpass property to reduce the high-frequency components and remove the DC term.
3. Fixed quantization levels: the masks should have an appropriate number of quantization levels to reduce the quantization error.

#### B. Problem Formulation

Our goal is to design an optimal mask, namely *OptMask*, that possesses the above-mentioned characteristics. Denote the desired random masks as $\mathbf{T}^{opt} \in \mathbb{C}^{N \times N \times L}$ with $|\mathbf{T}^{opt}| = 1$ for all pixels. Without loss of generality, we use the *l*-th mask $\mathbf{T}_l^{opt} \in \mathbb{C}^{N \times N}$ to represent $\mathbf{T}^{opt}$. Note that $\mathbf{T}_l^{opt}$ is a phase mask. Thus, it has a constant magnitude but varying phase angles. For simplicity, we define $\mathbf{T}_l^{opt} \coloneqq e^{i\breve{\psi}}$.

An intuitive solution for designing such masks is to design a filter with the required bandpass property and let it convolve with a white noise mask. However, the resulting mask may not be a phase mask and it also will not have fixed quantization levels. We need to include all these constraints in the design but it will make the optimization process non-convex. There is no off-the-shelf optimal filter design method [24], [25] for solving this unique problem.

For this reason, we propose a new two-stage optimization method for the design of the required mask. To begin with, we generate a phase-only white noise mask in the spectral domain $\mathcal{T} = e^{i\mathcal{Z}}$, where $\mathcal{Z} \sim U[0, 2\pi]$ represents that the phase angles are uniformly distributed over one trigonometric period. Then, we create a spectral filter $\boldsymbol{\Phi}$ to filter the unwanted frequency components. We then acquire the filtered masks $\overline{\mathcal{T}} = \mathcal{F}^{-1}\{\mathcal{T} \circ \boldsymbol{\Phi}\}$ in the spatial domain which will be used as a template for $\mathbf{T}^{opt}$. $\mathcal{F}^{-1}\{.\}$ represents the inverse Fourier transform. We make the optimal solution to be close to the template by minimizing the mean square error (MSE) between $\overline{\mathcal{T}}$ and $\mathbf{T}^{opt}$:

$$\breve{\psi} = \underset{\psi \in \mathbb{R}^{N \times N}}{\operatorname{argmin}} \left\| e^{i\psi} - \overline{\mathcal{T}} \right\|_2^2. \quad (5)$$

In this way, we ensure that the resulting $\breve{\psi}$ is a phase mask with the hope that it has a bandpass property similar to the template $\overline{\mathcal{T}}$. To ensure that the zero DC requirement is fulfilled (i.e., $\mathcal{F}\{e^{i\psi}\}(0) = \sum_{j,k} e^{i\psi_{j,k}} \approx \mathbb{E}[e^{i\psi}] = 0$), an additional constraint is added to the optimization problem as follows:

$$\breve{\psi} = \underset{\psi \in \mathbb{R}^{N \times N}}{\operatorname{argmin}} \left\| e^{i\psi} - \overline{\mathcal{T}} \right\|_2^2 \quad s.t. \quad \mathbf{1}^T vec(e^{i\psi}) = 0, \quad (6)$$

where $\mathbf{1} \in \mathbb{R}^{N^2 \times 1}$ and $vec(\cdot)$ denotes the all-one vector and vectorization operation, respectively. The augmented Lagrangian function of (6) can be formulated as:

$$\mathcal{L}(\psi, \gamma) = \min_{\psi \in \mathbb{R}^{N \times N}, \gamma \in \mathbb{C}} \frac{1}{2} \left\| e^{i\psi} - \overline{\mathcal{T}} \right\|_2^2 + \frac{\alpha}{2} \left| \mathbf{1}^T vec(e^{i\psi}) \right|^2 \quad (7)$$
$$+ Re < \mathbf{1}^T vec(e^{i\psi}), \gamma >,$$

where $\gamma$ is the Lagrangian multiplier for the equality constraint and $\alpha$ is a fixed constant for the penalty term. The resulting mask $\psi$ needs to be further quantized to allow it to be implemented by existing optical devices. The quantized mask should still follow the objective function set in (6).

#### C. Two-Stage Optimization Algorithm

We develop a novel two-stage optimization algorithm for solving the optimization problem in Section III-B. The details are described in the following sub-sections.

*1) Stage 1: Gradient Descent-Based Optimization*

Note that it is possible to derive a closed-form solution of $\psi^{(n)}$ at each iteration from (7). However, the optimization algorithm can empirically fail to converge to a local minimum if an inappropriate step size is selected. Considering that (7) is differentiable with respect to the decision variable $\psi$, we propose an efficient gradient descent-based algorithm for each pixel with an appropriately chosen step size. We first separate the Lagrangian function $\mathcal{L}(\psi, \gamma)$ into two parts: MSE and constraint.

   *a.* **MSE**: The derivative of the MSE term $\frac{1}{2}\left\|e^{i\psi} - \overline{\mathcal{T}}\right\|_2^2$ w.r.t. individual pixel $\psi_{j,k}^{(n)}$ at the $n$-th iteration is:

$$\frac{\partial \frac{1}{2}\left\|e^{i\psi} - \overline{\mathcal{T}}\right\|_2^2}{\partial \psi_{j,k}^{(n)}} = -\frac{\partial\left(\overline{\mathcal{T}}^H e^{i\psi^{(n)}} + \overline{\mathcal{T}}^T e^{-i\psi^{(n)}}\right)}{\partial \psi_{j,k}^{(n)}}$$

$$= -\frac{\partial\left(iIm\left(\overline{\mathcal{T}}^H e^{i\psi^{(n)}}\right)\right)}{\partial \psi_{j,k}^{(n)}} \quad (8)$$

$$= Re(\overline{\mathcal{T}}_{j,k})\sin(\psi_{j,k}^{(n)}) - Im(\overline{\mathcal{T}}_{j,k})\cos(\psi_{j,k}^{(n)})$$

$$:= \partial_\psi MSE_\mathcal{L}(\psi^{(n)}, \gamma^{(n)}).$$

   *b.* **Constraint**: The constraint term $\frac{\alpha}{2}\left|\mathbf{1}^T vec(e^{i\psi})\right|^2 + Re<\mathbf{1}^T vec(e^{i\psi}), \gamma>$ in (7) is equivalent to $\frac{\alpha}{2}\left|\mathbf{1}^T vec(e^{i\psi}) + \frac{\gamma}{\alpha}\right|^2 + C$, where $C$ is a constant unrelated to the derivative w.r.t. $\psi_{j,k}$. Hence,

$$\frac{\frac{\alpha}{2}\left|\mathbf{1}^T vec(e^{i\psi^{(n)}}) + \frac{\gamma^{(n)}}{\alpha}\right|^2 + C}{\partial \psi_{j,k}^{(n)}}$$

$$= \frac{\alpha\partial\left(\begin{array}{c}\sum_{j,k} e^{i\psi_{j,k}^{(n)}} \sum_{j,k} e^{-i\psi_{j,k}^{(n)}} + \frac{(\gamma^{(n)})^*}{\alpha}\sum_{j,k} e^{i\psi_{j,k}^{(n)}} \\ + \frac{\gamma^{(n)}}{\alpha}\sum_{j,k} e^{-i\psi_{j,k}^{(n)}}\end{array}\right)}{\partial \psi_{j,k}^{(n)}}$$

$$= \frac{\alpha\partial\left(\begin{array}{c}e^{i\psi_{j,k}^{(n)}}\sum_{[p,q]\neq[j,k]} e^{-i\psi_{p,q}^{(n)}} + e^{-i\psi_{j,k}^{(n)}}\sum_{[p,q]\neq[j,k]} e^{i\psi_{p,q}^{(n)}} \\ + \frac{(\gamma^{(n)})^*}{\alpha}e^{i\psi_{j,k}^{(n)}} + \frac{\gamma^{(n)}}{\alpha}e^{-i\psi_{j,k}^{(n)}}\end{array}\right)}{\partial \psi_{j,k}^{(n)}}$$

$$= -\left(Re(\gamma^{(n)})\sin\psi_{j,k}^{(n)} - Im(\gamma^{(n)})\cos\psi_{j,k}^{(n)}\right)$$

$$-\left(\sin\psi_{j,k}^{(n)} \alpha \sum_{[p,q]\neq[j,k]} \cos\psi_{p,q}^{(n)} - \cos\psi_{j,k}^{(n)} \alpha \sum_{[p,q]\neq[j,k]} \sin\psi_{p,q}^{(n)}\right)$$

$$:= \partial_\psi ConLoss_\mathcal{L}(\psi^{(n)}, \gamma^{(n)}) \quad (9)$$

Combining (8) and (9), the estimation at the $(n+1)$-th iteration is updated via the gradient descent formula:

$$\psi^{(n+1)} = \psi^{(n)} - \beta\partial_\psi\left(MSE_\mathcal{L}(\psi^{(n)} - \gamma^{(n)}) + ConLoss_\mathcal{L}(\psi^{(n)} - \gamma^{(n)})\right), \quad (10)$$

where $\beta$ denotes the step size of the gradient descent algorithm.

After acquiring $\psi^{(n+1)}$ with (10), the Lagrangian multiplier can be updated via,

$$\gamma^{(n+1)} = \gamma^{(n)} + \alpha\mathbf{1}^T vec\left(e^{i\psi^{(n+1)}}\right). \quad (11)$$

*2) Stage 2: Quick Search Pixel-Based Quantization*

After executing *Stage 1*, we acquire the optimal solution $\breve{\psi}$ which is a continuous variable with infinite precision. However, the commercial-grade SLMs have only 8-bit precision. We can only implement at most 256 phase changes with the phase-only SLMs. Therefore, it is necessary to quantize the continuous variable into a discrete one for successful implementation. The goal of the quantization problem is to assign each pixel of $\breve{\psi}$ to a specific finite value while minimizing the quantization error at the same time. To solve this problem, we propose a novel quick search pixel-based quantization method. Define an $M$-level codebook,

$$\mathcal{C} = \left[0, \frac{2\pi}{M}, 2\frac{2\pi}{M}, \dots, (M-1)\frac{2\pi}{M}\right], \quad (12)$$

with uniform spacing (usually, $M$ ranges from 2 to 16). We quantize $\breve{\psi}$ over $2\pi$ for two reasons: *(1)* usually SLMs are designed to modulate $0 - 2\pi$ phase shift; *(2)* quantization over one trigonometric period can prevent the phase wrapping problem. We quantize $\breve{\psi}$ pixel-by-pixel in numerical sequence from $[1,1]$ to $[N, N]$. At each iteration, we execute the quantization for one pixel.

When quantizing the $[j, k]$-th pixel of $\breve{\psi}$ at the $(j \times k)$-th iteration, we keep all the pixels except $\breve{\psi}_{j,k}$ to be the same as those at the $(j \times k - 1)$-th iteration and update $\breve{\psi}_{j,k}$ as follows:

$$\breve{\psi}_{p,q}^{(jk)} = \begin{cases} \breve{\psi}_{p,q}^{(j\times k-1)}, & [p,q] \neq [j,k] \\ \hat{\psi}_{j,k} \in \mathcal{C}, & [p,q] = [j,k] \end{cases} \forall p, q, \quad (13)$$

where $\hat{\psi}_{j,k}$ is obtained by projecting $\breve{\psi}^{(j\times k-1)}$ into the codebook $\mathcal{C}$ via a minimization operator $\mathcal{H}$ defined as follows:

$$\hat{\psi}_{j,k} = \mathcal{H}_{j,k}\left(\breve{\psi}^{(j\times k-1)}\right)$$
$$:= \min_{\breve{\psi}_{j,k}\in\mathcal{C}} \frac{1}{2}\left\|e^{i\breve{\psi}^{(j\times k-1)}} - \overline{\mathcal{T}}\right\|_2^2 \quad (14)$$
$$+ \frac{\alpha}{2}\left|\mathbf{1}^T vec\left(e^{i\breve{\psi}^{(j\times k-1)}}\right) + \frac{\breve{\gamma}}{\alpha}\right|^2$$

where $\breve{\gamma}$ denotes the optimal Lagrangian multiplier obtained from *Stage 1*. The quantized masks will still satisfy the conditions in (6). As can be seen, the optimization-based quantization method can quickly search for the optimal direction at each iteration and thus minimize the objective function. Besides, the proposed quantization method can satisfy the zero DC component property and bandpass attribute defined in Section III-B. In each loop, all pixels of the mask are processed. Empirically, we only need 2 loops to get a satisfactory performance. The whole algorithm is summarized below.

---

**Algorithm 1** Two-Stage Optimal Random Masks Design

**Input:** Lagrangian multiplier $\gamma = 0$, template $\bar{\mathcal{T}} \in \mathbb{C}^{N \times N}$, stopping threshold $\delta$, penalty constant $\alpha$, step size $\beta$, quantization loop number $G$, number of measurements $L$.

1: **for** $l \leftarrow 1$ to $L$ **do**
2:    **Stage 1**: *Gradient Descent-Based Optimization*
3:    **while** $\frac{\|\psi^{(n+1)} - \psi^{(n)}\|_2^2}{N^2} \geq \delta$ **do**
4:      Update $\psi^{(n+1)}$      ▷ Eq. (10)
5:      $\gamma^{(n+1)} = \gamma^{(n)} + \alpha 1^T vec\left(e^{i\psi^{(n+1)}}\right)$
6:    **end while**
7:    **Stage 2**: *Quick Search Pixel-Based Quantization*
8:    Create an M-level codebook $\mathcal{C}$     ▷ Eq. (12)
9:    **for** $g \leftarrow 1$ to $G$ **do**
10:      **for** $[j, k] \leftarrow [1,1]$ to $[N, N]$ **do**
11:        $\breve{\psi}_{p,q}^{(jk)} = \begin{cases} \breve{\psi}_{p,q}^{(j \times k - 1)}, & [p,q] \neq [j,k] \\ \mathcal{H}_{j,k}(\breve{\psi}^{(j \times k - 1)}), & [p,q] = [j,k] \end{cases}$
                               $\forall p, q$   ▷ Eqs. (13, 14)
12:      **end for**
13:    **end for**
14:    $\mathbf{T}_l^{opt} = e^{i\breve{\psi}}$
15: **end for**

**Output:** *OptMask* $\mathbf{T}^{opt}$

---

### D. Reconstruction Algorithm

Recovering the complex-valued images from the phaseless intensity measurements is known to be a nonconvex optimization problem, where the forward model is the magnitude square of the masked image's Fourier transform. Regularization terms subject to the image prior should be added to the optimization process to reduce the measurement noise while preserving important details such as edges and textures. In this study, we adopted total-variation (TV) regularization to smooth out the noise while preserving the image edges. Combined with the Maximum A-Posterior (MAP) loss function, we adopt the TV-MAP algorithm [30] as the CDP phase retrieval algorithm in this study:

$$\min_{\mathbf{x} \in \mathbb{C}^n} \alpha \nabla \|\mathbf{x}\|_1 + \frac{1}{2} \sum_i (|x_i|^2 + \mathcal{X}_i \log |x|_i), \quad (15)$$
$$s.t. \ \mathcal{X} = |\mathcal{F}(\mathbf{T} \circ \mathbf{x})|^2,$$

where $\nabla$ is the gradient operator, $\|.\|_1$ denotes the $\ell_1$ norm. Based on the assumption that images with spurious details have higher total variation, the TV regularization term is implemented in (15) for removing the noise, which appears in the first term of the equation. Reducing the total variation of the measurement can remove unwanted details. On the other hand, the MAP data fidelity term (the second term in (15)), acting as the loss function in (15), evaluates the performance of the estimation. Minimizing these two terms can obtain an estimate that is close to the original $\mathbf{x}$ while reducing the spurious details because of the noise. Furthermore, the optimal solution allows discontinuities along the curves; therefore, edges can be preserved in the restored image. The regularization strength is controlled by the constant $\alpha$, which is usually determined empirically according to the complexity of the desired signal. (15) can be solved via the Alternating Direction Method of Multipliers (ADMM) algorithm [31].

## IV. ANALYSIS OF THE PROPOSED MASKING SCHEME

In this section, we analyze some crucial factors and determine the appropriate parameters of the proposed *OptMask* in the simulation environment. To quantitatively measure the effects brought by different factors, we define a parameter $\eta$ to indicate the high-frequency energy of the masks:

$$\eta(\mathbf{\Psi}) = \frac{\sum_{j,k} \mathbf{\Psi}_{[j,k \geq 0.8N]}}{\sum_{j,k} \mathbf{\Psi}_{j,k}} \quad (16)$$

where $\mathbf{\Psi} = |DFT(\mathbf{T}^{opt})|^2$. Basically, if a mask has a larger $\eta$, the resulting Fourier intensity measurements will have more high-frequency data beyond the Nyquist rate of the system. Then, there will be more high-frequency data removed in the phase retrieval computation.

As to the other parameters, the size of the masks was $256 \times 256$. The size of the oversampling Fourier intensity measurements was $762 \times 762$ (the same size in the optical experimental setup). We set the pixel size as $8\mu m$ and the wavelength $\lambda$ as $632.8nm$ (assuming HeNe laser is used). In addition, we simulated the "dead zone areas" (small gaps) between consecutive cell units of the SLM device [26]. There is no phase modulation in these areas. It introduces a small bias to the mask in the spatial domain. The penalty constant $\alpha$ in (7) and step size $\beta$ in (10) were empirically set as $10^{-4}$ and $0.2$, respectively. The maximum number of iterations for *Stage 1* of the 2-stage algorithm is 300 and the stopping threshold was $10^{-7}$.

### A. Spectral Filter

As mentioned in Section III-B, we use a template $\bar{\mathcal{T}}$ to guide *OptMask* to have the desired bandpass frequency response. The template $\bar{\mathcal{T}}$ was generated by applying a spectral filter $\mathbf{\Phi}$ with the desired frequency response to a random mask. The main parameter of $\mathbf{\Phi}$ is its passband frequencies. They have important consequences on the randomness and high-frequency components of the mask, as will be shown below. To let us focus on the passband frequencies, we implemented $\mathbf{\Phi}$ with an ideal filter [21], which is flat within the passband and has a sharp transition band, making it a good candidate for spectral filtering. An example of the process to acquire the template $\overline{\mathcal{T}}$ through an ideal filter $\boldsymbol{\Phi}$ is shown in Fig. 3.

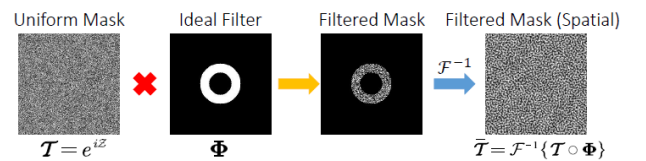

Fig. 3: Illustration of generating the template $\bar{\mathcal{T}}$ via an ideal filter.

*B. Passband Frequency Analysis*

In this section, we investigated the effect of the passband frequency of $\boldsymbol{\Phi}$ on *OptMask*. Note that the true intensity image is the convolution integral of the mask's Fourier spectrum and the target image's Fourier spectrum. If the spectrum of the mask has a broader passband, the convolution kernel will cover a larger area in the Fourier plane but have a lower gain. For most natural images, in high-frequency regions, significant Fourier coefficients are often clustered along the horizontal and vertical axes. The convolution of such a spectrum with a large but low gain kernel will not generate much significant data in high-frequency regions. On the contrary, for a mask with a narrower passband, the convolution kernel will cover a smaller area in the Fourier plane but have a larger gain. It can generate significant high-frequency data when the kernel, with a higher gain, convolves with the large Fourier coefficients of the image in the high-frequency regions (although not many). To illustrate this, we generated $L = 10$ masks for random masks of different bandwidths and passband frequencies. We multiplied these masks with an image and computed the Fourier intensity of the masked images using DFT. The resulting Fourier intensities using different masks are shown in Fig. 4. Here, we borrow the operator $\eta$ defined in (16) to evaluate the high-frequency contents of these Fourier intensities. Note that these Fourier intensities are not the same as those we measured in the experiment since they are computed using DFT (the real ones are obtained via the Fourier transform optically). They do not have data outside the Nyquist frequency. However, their high-frequency energy close to the Nyquist frequency can be a good indicator of the high-frequency energy of the real measurements. So we computed the average $\overline{\eta} = \frac{1}{L}\sum_{l=1}^{L}\eta_l(\boldsymbol{\Psi}_l)$ of these intensity measurements. The results are also shown in Fig. 4. For all optimal masks, the first value refers to the lower cut-off frequency and the second value denotes the upper cut-off frequency. As can be seen in the first column, two bandpass random masks have different lower cut-off frequencies ($\pi/4$ and $\pi/5$) but the same upper cut-off frequency $\pi/2$. The Fourier intensity of that with a lower cut-off frequency $\pi/5$ (thus broader passband) has a smaller $\eta$, i.e., less significant high-frequency content. It is the same for the intensity images in the second column. The Fourier intensity of that with a lower cut-off frequency $\pi/5$ (thus broader passband) again has a smaller $\eta$. These results verify our argument above. Thus, if we just consider the high-frequency content of the intensity measurement, the filter $\boldsymbol{\Phi}$ should have a broader passband. However, we will show in Section IV-D that we also need to consider the randomness of the mask. It demands an optimal design to balance the two requirements.

*C. Quantization Level*

As derived in Section III-C, the acquired random masks after *Stage 1* of the proposed mask design algorithm are continuous variables. It is necessary to quantize the random masks for loading into hardware devices, such as SLM. We, therefore, developed the *quick-search pixel-based quantization* method in *Stage 2* for efficient quantization. In this section, we analyze the effect of the number of quantization levels on the high-frequency content of the mask. Although we design the optimal mask with no high-frequency contents in *Stage 1*, the high-frequency data can be re-introduced to the mask after it is quantized. To illustrate this, an experiment was carried out to evaluate the high-frequency content of *OptMask* when it is quantized with different numbers of levels. The results are presented in Fig. 5. As can be seen in Fig. 5(a), the high-frequency energy parameter $\eta$ (defined in (16)) of the 2-level (binary) and 4-level masks are much higher than the unquantized mask. It is because the quantization introduces many sharp changes between adjacent pixels if there are not enough quantization levels. On the other hand, $\eta$ decreases with the growth of the quantization levels. As can be seen, the 16-level and 32-level masks have $\eta$ comparable to the unquantized mask. In practice, the gray-scale phase mismatch due to the imperfections of SLM devices can bring quantization errors. Some pixel values cannot be correctly implemented because of the non-linear input-output mappings of the devices. Therefore, as 16-level masks and 32-level masks have similar $\eta$, we adopted the 16-level masks to reduce our reliance on the accuracy of SLM. Fig. 5(b) shows a comparison of the power spectra (along the x-axis) of different masks. The difference in the high-frequency power between the binary mask and the 16-level mask can be up to 10 dB.

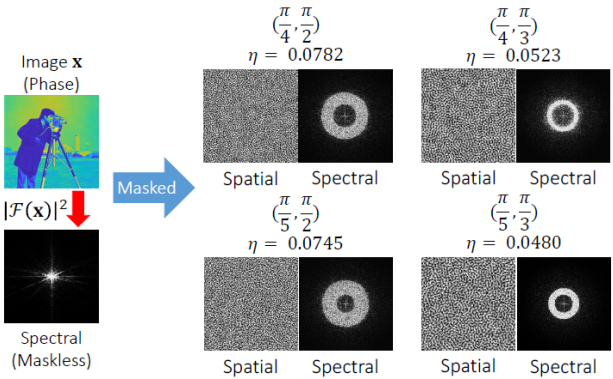

Fig. 4: (a) Visualization of the optimal mask of different bandwidths and passband frequencies in the spatial and spectral domains.

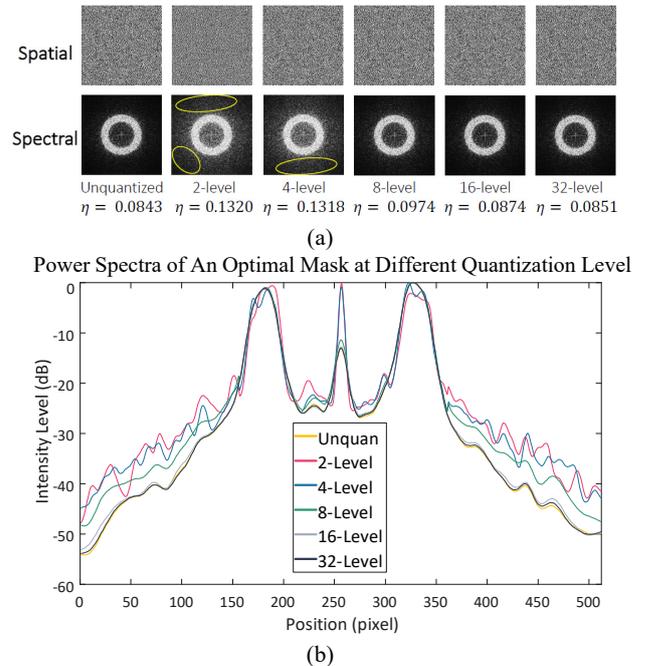

Fig. 5: (a) Spatial and spectral domains of the bandpass optimal mask ($\pi/3, \pi/2$) at different quantization levels. (b) Mean power Spectra (in log scale) of the optimal masks in (a) along the x-axis.

*D. Randomness*

The randomness of a random mask plays an important role in the reconstruction performance. Intuitively, the more random the mask, the better the reconstruction performance of the CDP phase retrieval algorithm. It is because higher randomness brings more information into the measurement process that can reduce the ill-posedness of the phase retrieval problem [9]. In the last decade, various randomness measurement methods based on the statistical information of digital images were proposed [27], [28]. Shannon entropy measure [29], defined as $S(X) = E[-\log p(X)]$ for a discrete random variable $X$, is widely used for measuring randomness since it provides the quantitative description of uncertainty. The larger $S(X)$, the more uncertain the signal $X$ is, thus meaning higher randomness.

To investigate the randomness of *OptMask* under different settings, we adopted the local Shannon entropy measure, a generalization of the Shannon entropy, to quantify randomness [27]. The local Shannon entropy splits an image $X$ into $k$ non-overlapping blocks and computes the average Shannon entropy $\overline{S}_k(X) = \frac{1}{k}\sum_{i=1}^{k} S(X_i)$ over $k$ blocks, where $X_i$ denotes the $i$-th block. We generated $L = 10$ masks for each kind of random mask and calculated $\overline{S}$. We used 16-level optimal random masks and 32×32 image blocks. For the green noise mask, we used 16×16 image blocks. The visualization of different masks is shown in Fig. 6. As can be seen, the entropy of the green noise mask is much smaller than the optimal masks. It is because the theoretical maximum entropy of a signal is closely related to the number of intensity levels. The binary green noise masks can only have uncertainty at most $S = \log_2 2 = 1$ while multiple-level optimal masks (16-bit in the figure) have $En = \log_2 16 = 4$. The multiple-level representation of the mask thus increases the randomness and helps improve the reconstruction performance. On the other hand, the uncertainty decreases with the decrease in the lower cut-off frequency. It is because pixels need to be clustered to achieve the low-frequency response of the mask. The lower the frequency, the larger the clusters is, which in turn lowers the randomness of the mask. Although using the optimal masks with lower cut-off frequencies (or broader passband) will incur less high-frequency data in the intensity measurements (as shown in the analysis in Section IV-B), we choose the optimal mask with cut-off frequencies $(\pi/5, \pi/3)$ as the final design. It is based on a holistic consideration of both the randomness and high-frequency energy of the mask.

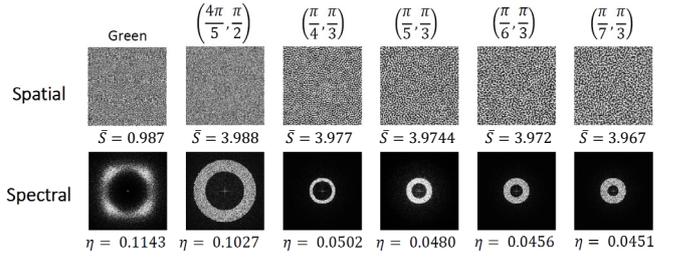

Fig. 6: Randomness (quantified by average entropy) of different kinds of random masks.

## V. EXPERIMENT

*A. Experimental Setup*

**Optical Setup**: We evaluated the proposed *OptMask* on a practical optical system which is shown in Fig. 7. It contains a Thorlabs $10mW$ HeNe laser with wavelength $\lambda = 632.8nm$; and a 12-bit 1920×1200 Thorlabs Kiralux CMOS camera with pixel pitch $5.86\mu m$. We used a 1920 × 1080 Holoeye Pluto phase-only SLM with a pixel pitch $\delta_{SLM} = 8\mu m$ to generate the phase-only random masks in the experiments. We only utilized the central 256 × 256 SLM pixels in all experiments. The resulting Fourier intensity measurements were captured by the imaging sensors. The size of the acquired measurement is $\frac{\lambda f}{d \delta_{SLM}} = 762 \times 762$, where $d$ and $f$ denote the pixel pitch of the camera and the focal length of the lens before the camera, respectively.

**Target images**: We used two kinds of target images for experimental evaluation: complex-valued images (phase-only) and real-valued images. For complex-valued phase-only images, we pre-multiplied the phase-only images with the masks and loaded them into the SLM. In this case, the SLM was not only for implementing the phase mask but also for the phase-only images. For the real-valued images, we adopted a USAF chart (Newport Inc.) as the testing object, as shown in Fig. 11.

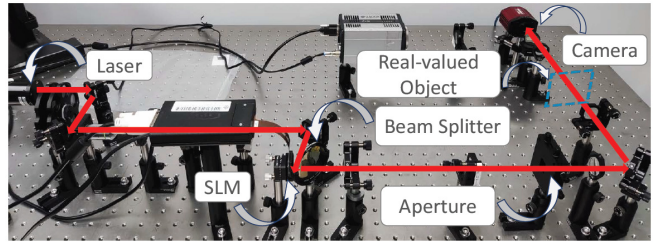

Fig. 7: The hardware setup of the SLM-based CDP phase retrieval system. It implements the optical path as shown in Fig. 1.

**Error Criteria**: We used two criteria to evaluate the reconstruction performance of the image phases: peak signal-to-noise ratio (PSNR) and structural similarity index (SSIM). They are defined as follows:

$$PSNR_{phase}(\mathbf{x}, \tilde{\mathbf{x}}) = 10 \log_{10} \sum_{j=0}^{M-1} \frac{2\pi}{MSE_{phase}(\mathbf{x}, \tilde{\mathbf{x}})}, \quad (17)$$

$$SSIM_{phase}(\mathbf{x}, \tilde{\mathbf{x}}) = SSIM(\angle \mathbf{x}, \angle \tilde{\mathbf{x}} - \theta),$$

where $MSE_{phase}(\mathbf{x}, \tilde{\mathbf{x}}) = \frac{1}{MN}\sum_{i=0}^{N-1}\sum_{j=0}^{N-1}(\angle x_{i,j} - \angle \tilde{x}_{i,j} - \theta)^2$. $\mathbf{x}$ and $\tilde{\mathbf{x}}$ represent the original and reconstructed phase images, respectively. Both $\angle \mathbf{x}$ and $\angle \tilde{\mathbf{x}}$ are unwrapped to compare the true phase difference. Besides, $\pi$ phase shift is added to $\angle \mathbf{x}$ and $\angle \tilde{\mathbf{x}}$ to ensure non-negative values when computing $PSNR_{phase}$. A global phase shift term $\theta$ due to ambiguity is removed to avoid the $MSE_{phase}$ value being amplified.

### B. Experimental Results

#### 1) Number of Measurements

Intuitively, better reconstruction performance can be achieved by using a larger number of measurements for reconstruction since more information is introduced. However, using a large number of measurements would lengthen the image acquisition time and require the testing object to be static for a long period. It affects the reconstruction performance of dynamic objects. To investigate the appropriate numbers of measurements, we used a natural image named "crowd" as the ground truth to investigate the effects of the number of measurements. We collected 4 masked Fourier intensity measurements via *OptMask* and reconstructed the images with different numbers of measurements with the same hyperparameters. The results are shown in Fig. 8. As can be seen, although it is possible to reconstruct the global profile of the ground truth with only one measurement, the details of the images are nearly lost. While the PSNR and SSIM increase with the growth of the number of measurements, there is no substantial difference between 3 masks and 4 masks cases because of sufficient information. Therefore, we adopted 3 measurements for the following experiments to balance performance and efficiency.

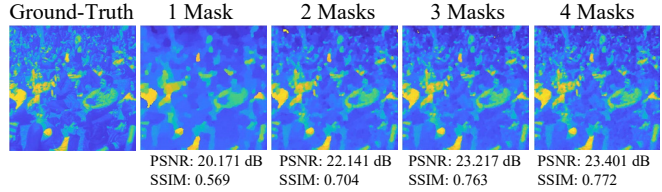

Fig. 8: The reconstructed images (phase part) with different numbers of Fourier intensity measurements. The target image is masked with *OptMask*.

#### 2) Masks with Different Cut-off Frequencies

As analyzed in Section IV-D, the cut-off frequencies of the optimal masks can affect their high-frequency contents and randomness. They thus have an impact on the reconstruction performance. To verify this, we conducted experiments with masks of different cut-off frequencies. Two natural images namely "crowd" and "Triumphal Arch" were used as the ground truth. The compared masks include the binary green noise masks and three 16-level optimal masks of different cutoff frequencies. We collected 3 Fourier intensity measurements for each kind of mask and reconstructed the images with the same hyperparameters. The results are shown in Fig. 9. As can be seen, although *OptMask* with cut-off frequencies $(4\pi/5, \pi/2)$ and the green noise mask have a similar amount of high-frequency information $\eta$ (see Fig. 6), the reconstruction performance with the multiple-level optimal mask is much better than that of the binary green noise mask. It is due to the large difference in their entropy (and hence the randomness). Comparing the multiple-level *OptMasks*, the one with the cut-off frequencies $(\pi/5, \pi/3)$ gives the best performance, as shown in Fig. 9. It strikes a balance in the entropy and $\eta$ value, as shown in Fig. 6. The above result verifies our choice of using that mask as our final design.

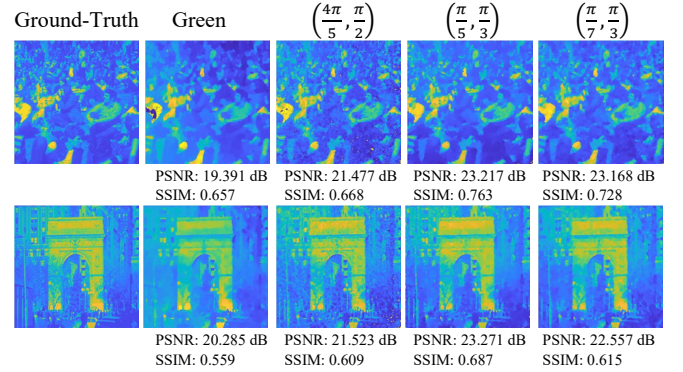

Fig. 9: The reconstructed images (phase part) with masks of different cut-off frequencies.

#### 3) Comparison with Other Masking Schemes

We compared the proposed *OptMask* with four masking schemes: *(1)* 4-level white noise masks (denoted as White4) [9]; *(2)* 16-level white noise masks (denoted as White16) [9]; *(3)* binary green noise masks (denoted as Green) [18]; *(4)* correlation-based coded apertures (denoted as SR-CDP) [17]; and *(5)* admissible coded apertures (denoted as PR-CDP) [16]. For *OptMask*, we adopted the one with the cut-off frequency $(\pi/5, \pi/3)$ (as mentioned in Section IV-D). For Green, 2-level masks with a discrete uniform distribution $d \in \{-1,1\}$ were used, as the original setting. For White-4, 4-level masks with a discrete uniform distribution $d \in \{-1,1,-j,j\}$ were used. For White-16, 16-level masks with a discrete uniform distribution $d \in [-\pi, \pi]$ (same codebook of *OptMask*) were used. For PR-CDP and SR-CDP, we extend their original settings to let them use the same 16-level uniformly distributed codebook as *OptMask*. The above allows us to compare White-16, PR-CDP, PR-CDP, and the proposed *OptMask* on the same ground. It is worth noting that, except *OptMask*, all the other masks are designed directly with discrete random variables and no quantization process is involved. Some examples of the compared masks and their corresponding Fourier intensities are shown in Fig. 10. It can be seen that the proposed *OptMask* has the smallest $\eta$. It is one of the key factors that lead to improved performance, as we will demonstrate in the following sections.

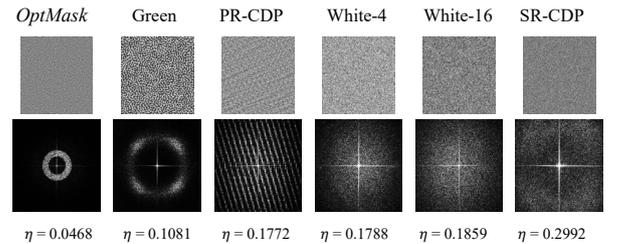

Fig. 10: The spatial domain (phase part) of different kinds of masks and their corresponding Fourier intensity measurements. $\eta$ is computed through an average of 4 measurements.

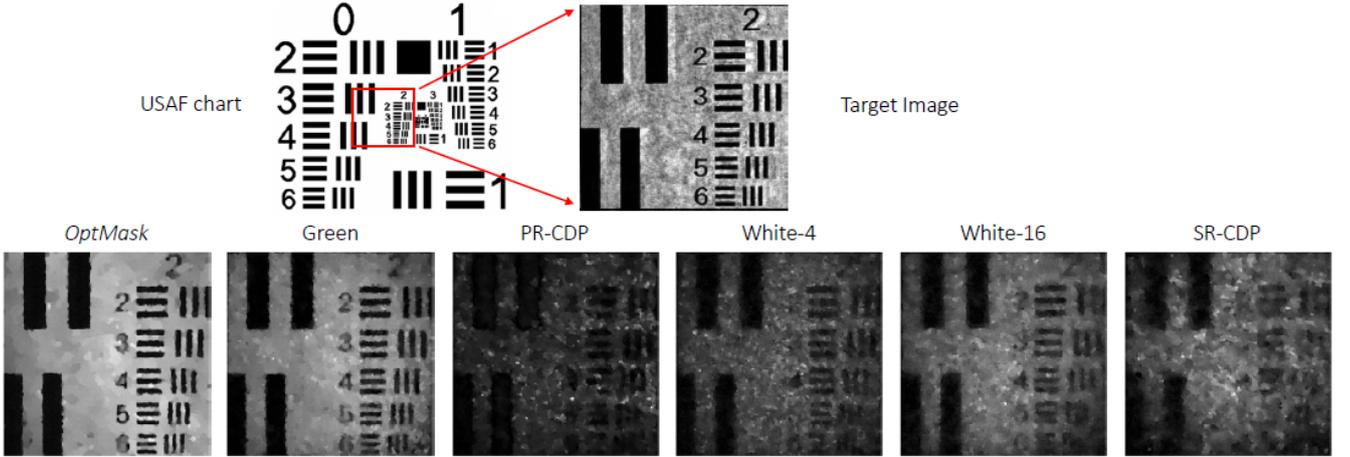

Fig. 11: Experimental results of different phase retrieval masking schemes on a USAF chart. The first row shows the whole USAF chart image and the target image captured by the imaging sensor (zoom-in red-rectangle regions). The second row presents the reconstructed images through different masking schemes.

• **Real-Valued Images**: As mentioned above, we placed a USAF chart on the object plane of the experiment setup as the target object. A picture of the target image is shown in Fig. 11. The chart only allows the projected light to go through its chart's holes (dark areas in the figure). Thus, the resulting target images have sharp amplitude changes along the boundaries of the holes. The regions that the light can pass through should have the same phase thus we treat the target image as a real-valued binary image. Note that for each masking scheme, we fine-tuned the parameters of the reconstruction algorithm to obtain the best visual effect. We ran 100 iterations of the reconstruction algorithm and three trials for all situations. For the experiment, we captured three measurements via the imaging sensor for each kind of mask.

The qualitative results are shown in Fig. 11. Since the USAF chart is a binary testing object, we demonstrated the visual comparisons in grayscale images. It can be seen that the estimated images through different kinds of binary masks have clear differences. Even though the general profile of the target image can be identified by all masking schemes, the quality of the reconstructed image has a considerable difference. The reconstructed image resulting from using *OptMask* remains sharp at the edge regions and relatively smooth in the background areas. The reconstructed image via the binary Green noise masks is a bit blurry although it has a relatively smooth background. For other kinds of masks, it is difficult to recognize the details of the images. The above experimental results show that the proposed *OptMask* outperforms all state-of-the-art masking schemes qualitatively when applied to practical phase retrieval system testing with a real object.

• **Complex-Valued Images**: We used the phase-only images as the testing objects as described in Section V-A. We chose 4 natural images and 4 standard testing images. Specifically, the standard testing images include an optical vortex image and a checkerboard pattern image from the SLM built-in software, and 2 cell images from Wikipedia that follow the public domain licenses. We ran 200 iterations of the reconstruction algorithm and three trials for all situations. We chose the results with the highest PSNR value.

The quantitative and qualitative results are shown in TABLE I and Fig. 12, respectively. As presented in Table I, the proposed *OptMask* consistently outperforms all compared masking schemes in PSNR and SSIM. For natural images, *OptMask* has PSNR and SSIM gains of at least $2.605dB$ and 0.0.073, respectively. As shown in Fig. 12, the proposed *OptMask* can reconstruct images with the most similar contours as the ground truths and preserve more details compared with using other masking schemes. Specifically, the results via

TABLE I: Quantitative comparison (PSNR/SSIM) with the state-of-the-art masking schemes for phase retrieval on natural and standard testing images. The corresponding qualitative results are shown in Fig. 12. Best performances are marked in bold and the second best performances are colored in green. The Fourier intensity measurements are collected with the optical system presented in Fig. 7.

| Masks / Images | *OptMask* (Ours) PSNR ↑ | *OptMask* (Ours) SSIM ↑ | Green PSNR ↑ | Green SSIM ↑ | PR-CDP PSNR ↑ | PR-CDP SSIM ↑ | White-4 PSNR ↑ | White-4 SSIM ↑ | White-16 PSNR ↑ | White-16 SSIM ↑ | SR-CDP PSNR ↑ | SR-CDP SSIM ↑ |
|---|---|---|---|---|---|---|---|---|---|---|---|---|
| Natural Images | | | | | | | | | | | | |
| Triumphal Arch | **23.271** | **0.667** | 20.285 | 0.559 | 16.312 | 0.480 | 14.566 | 0.471 | 16.404 | 0.493 | 15.705 | 0.426 |
| Tree | **18.986** | **0.638** | 17.197 | 0.434 | 13.214 | 0.357 | 16.360 | 0.495 | 16.858 | 0.511 | 12.606 | 0.303 |
| Crowd | **23.217** | **0.763** | 19.391 | 0.657 | 14.784 | 0.549 | 16.778 | 0.528 | 18.127 | 0.481 | 13.664 | 0.476 |
| Sculpture | **21.751** | **0.703** | 19.126 | 0.533 | 13.991 | 0.429 | 15.161 | 0.542 | 16.419 | 0.549 | 13.269 | 0.390 |
| Standard Testing Images | | | | | | | | | | | | |
| Vortex | **18.344** | **0.826** | 15.604 | 0.732 | 14.694 | 0.679 | 15.111 | 0.597 | 15.346 | 0.631 | 13.191 | 0.548 |
| Cell | **22.093** | **0.745** | 19.734 | 0.614 | 17.144 | 0.539 | 19.138 | 0.633 | 18.339 | 0.584 | 15.952 | 0.504 |
| Cell Cluster | **22.634** | **0.811** | 18.288 | 0.744 | 13.131 | 0.578 | 15.001 | 0.564 | 16.478 | 0.555 | 12.474 | 0.494 |
| Checkerboard | **22.159** | **0.859** | 20.561 | 0.752 | 13.966 | 0.533 | 13.803 | 0.591 | 14.201 | 0.584 | 11.865 | 0.465 |

*OptMask* can successfully reconstruct the complex details and textures of the "tree" and "crowd" images. As for the reconstructed images via other masking schemes, the textures are nearly lost. Besides, the phase jumping errors appear in regions near 0 (dark blue) or $2\pi$ (light yellow). For the standard testing images, the proposed *OptMask* can achieve PSNR and SSIM gains of at least $2.105dB$ and $0.082$, respectively, compared with other masking schemes. It can be seen in Fig. 12 that the reconstructed images via *OptMask* have better quality than those using other masks. *OptMask* cannot only recover the contours and detailed textures of the ground truth but also avoid the phase jumping error compared with other masking schemes. The above experimental results show that the proposed *OptMask* outperforms all state-of-the-art masking schemes quantitatively and qualitatively when used in a practical phase retrieval system.

## VI. CONCLUSION

This paper proposed a novel multiple-level green noise phase mask and its design method for practical coded diffraction pattern (CDP) phase retrieval. The proposed *OptMask* has very few high-frequency components while maintaining the required randomness. They are important to the performance of the CDP phase retrieval. *OptMask* is designed based on a two-stage optimization algorithm that allows the flexibility to adjust the cut-off frequencies and quantization levels. We have demonstrated how we optimally select the cut-off frequencies and quantization levels of *OptMask* to minimize its high-frequency content and maximize its randomness. We demonstrated the performance of *OptMask* by applying it to a practical CDP phase retrieval system. The results showed that the quality of the amplitude and phase images retrieved using the proposed *OptMask* significantly outperformed the traditional CDP masking schemes both qualitatively and quantitatively.

**Appendix: Calibration**

Since the phase objects and the coded apertures in our experiments are generated by the SLM of the system, we need to carefully calibrate the SLM such that we can generate the masked target images and quantitatively evaluate the fidelity of our reconstructed images. We calibrated the phase-only SLM using the interferometric configuration [32]. The interferometric approach can be used for recording and reconstructing amplitude and phase by mathematical techniques. For example, under the same direction of polarization, the interference fringes of two plane waves are described as:

$$I(x,y) = I_1(x,y) + I_2(x,y) + 2\sqrt{I_1(x,y)I_2(x,y)}(\cos\Delta), \quad (a.1)$$

where $I_1(x,y)$ and $I_2(x,y)$ denote the intensities on the measurement plane of two light beams, $I(x,y)$ refers to the interference pattern and $\Delta$ represents the phase difference between two beams. The interference pattern can be realized with different parts of SLM's active areas.

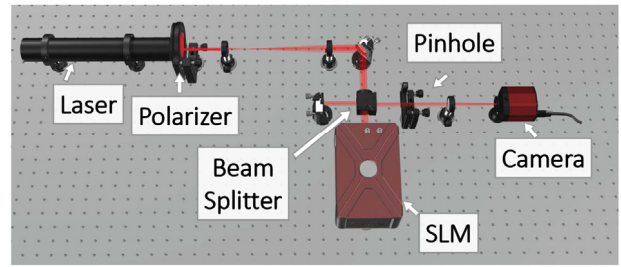

Fig. 13: Configuration of the phase calibration with a reflective SLM.

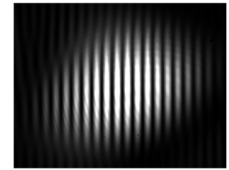

(a)

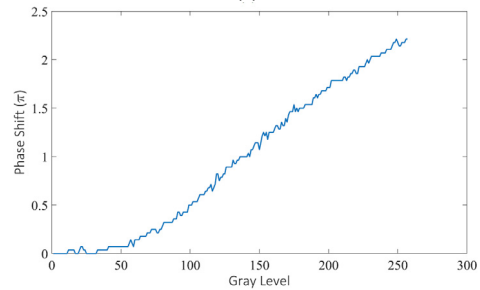

(b)

Fig. 14: (a) An example of the interference pattern. (b) Experimentally calibrated intensity modulation curve in the grayscale range of $[0, 255]$ for the SLM.

We applied the self-reference approach that splits the SLM plane into two zones (half-half) to behave as a phase-shift zone and a reference zone, respectively [33]. The optical path of the calibration system is shown in Fig. 13. Specifically, we uploaded a series of patterns with a constant grayscale value (from 0 to 255) on the modulated area. Then the wavefront transmitting the region was adjusted with a specific phase shift according to the assigned grayscale on the SLM areas. At the same time, the addressed value on the reference regions remained unchanged. To create the interference patterns effectively, we placed a pinhole board behind the SLM device, which is an opaque card with two transparent pinholes that only allow the light beam to pass through the pinholes. We carefully designed the pinhole board such that each pinhole locates at the modulated zone and reference zone, respectively. An example of the interference pattern is shown in Fig. 14(a) and the variation of phase modulation with 8-bit grayscale values is presented in Fig. 14(b). As can be seen, the effective phase modulation range exceeds $2\pi$ for the Holoeye SLM in the wavelength of $632nm$. To prevent phase wrapping, we only utilize the phase shift with the range of $[0, 2\pi]$ that corresponds to $[30, 230]$ of the grayscale values. Then we input the corresponding phase shift of each grayscale value into a look-up-table file and upload it to the SLM.

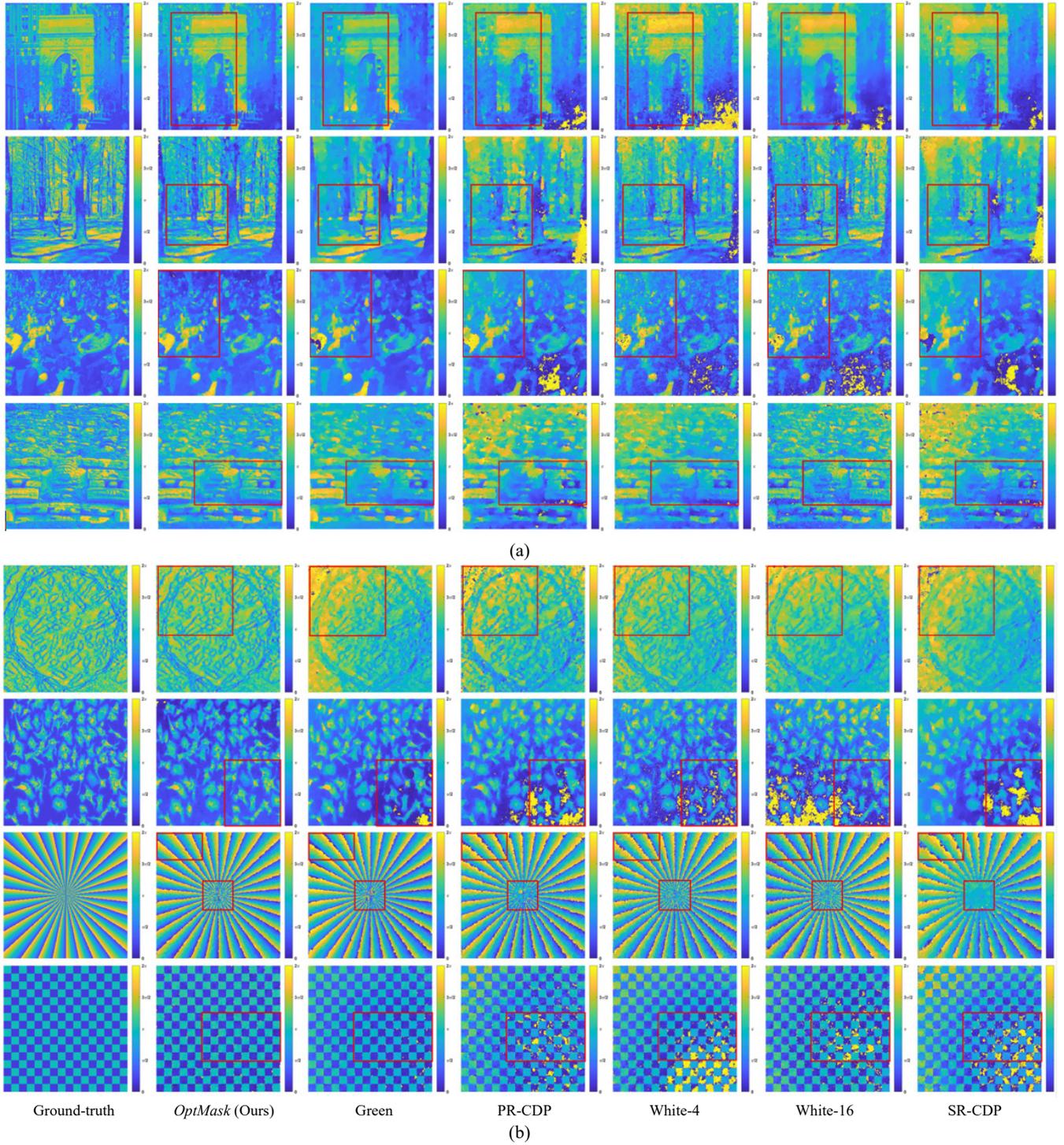

Fig. 12: Experimental results of different phase retrieval masking schemes on (a) natural images, and (b) standard testing images. The first column shows the phase part of the ground truth images with color bars. The other columns show the reconstructed images using different masking schemes. The colormap of all columns ranges from 0 to $2\pi$. The regions inside the red rectangular boxes represent the most obvious visual differences between different masking schemes. The quantitative results (PSNR/SSIM) of the reconstructed images are shown in TABLE I.

## References

[1] Y. Shechtman, Y. C. Eldar, O. Cohen, H. Chapman, J. Miao, and M. Segev, "Phase retrieval with application to optical imaging: A contemporary overview," *IEEE Signal Processing Magazine*, vol. 32, pp. 87–109, 2015.

[2] R. W. Gerchberg, "A practical algorithm for the determination of phase from image and diffraction plane pictures," *Optik*, vol. 35, pp. 237–246, 1972.

[3] J. R. Fienup, "Phase retrieval algorithms: a comparison," *Appl. Opt.*, vol. 21, no. 15, pp. 2758–2769, Aug 1982.

[4] J. Miao, P. Charalambous, J. Kirz, and D. Sayre, "Extending the methodology of x-ray crystallography to allow imaging of micrometresized non-crystalline specimens," *Nature*, vol. 400, pp. 342–344, 1999.

[5] J. M. Rodenburg, "Ptychography and related diffractive imaging methods," *Advances in Imaging and Electron Physics*, vol. 150, pp. 87–184, 2008.

[6] Y. Xu, Q. Ye, A. Hoorfar, and G. Meng, "Extrapolative phase retrieval based on a hybrid of phasecut and alternating projection techniques," *Optics and Lasers in Engineering*, vol. 121, pp. 96–103, 2019.

[7] C. Shen, C. Guo, J. Tan, S. Liu, and Z. Liu, "Complex amplitude reconstruction by iterative amplitude-phase retrieval algorithm with reference," *Optics and Lasers in Engineering*, vol. 105, pp. 54–59, 2018.

[8] C. Guo, C. Shen, J. Tan, X. Bao, S. Liu, and Z. Liu, "A robust multiimage phase retrieval," *Optics and Lasers in Engineering*, vol. 101, pp. 16–22, 2018.

[9] E. J. Candes, X. Li, and M. Soltanolkotabi, "Phase retrieval from coded diffraction patterns," *Applied and Computational Harmonic Analysis*, vol. 39, no. 2, pp. 277–299, 2015.

[10] ——, "Phase retrieval via wirtinger flow: Theory and algorithms," *IEEE Transactions on Information Theory*, vol. 61, no. 4, pp. 1985–2007, 2015.

[11] C. Falldorf, M. Agour, C. v Kopylow, and R. Bergmann, "Phase retrieval by means of a spatial light modulator in the fourier domain of an imaging system," *Applied optics*, vol. 49 10, pp. 1826–30, 2010.

[12] R. Horisaki, Y. Ogura, M. Aino, and J. Tanida, "Single-shot phase imaging with a coded aperture," *Optics letters*, vol. 39 22, pp. 6466–9, 2014.

[13] R. Horisaki, R. Egami, and J. Tanida, "Experimental demonstration of single-shot phase imaging with a coded aperture," *Optics express*, vol. 23 22, pp. 28691–7, 2015.

[14] C. Zheng, R. Zhou, C. Kuang, G. Zhao, Z. Yaqoob, and P. So, "Digital micromirror device-based common-path quantitative phase imaging," *Optics Letters*, vol. 42 7, pp. 1448–1451, 2017.

[15] V. Boominathan, J. K. Adams, J. T. Robinson, and A. Veeraraghavan, "Phlatcam: Designed phase-mask based thin lensless camera," *IEEE Transactions on Pattern Analysis and Machine Intelligence*, vol. 42, no. 7, pp. 1618–1629, 2020.

[16] A. Guerrero, S. Pinilla, and H. Arguello, "Phase recovery guarantees from designed coded diffraction patterns in optical imaging," *IEEE Transactions on Image Processing*, vol. 29, pp. 5687–5697, 2020.

[17] J. Bacca, S. Pinilla, and H. Arguello, "Super-resolution phase retrieval from designed coded diffraction patterns," *IEEE Transactions on Image Processing*, vol. 29, pp. 2598–2609, 2020.

[18] Q. Ye, Y.-H. Chan, M. G. Somekh, and D. P. Lun, "Robust phase retrieval with green noise binary masks," *Optics and Lasers in Engineering*, vol. 149, p. 106808, 2022.

[19] Y.-H. Fung and Y.-H. Chan, "Green noise digital halftoning with multiscale error diffusion," *IEEE Transactions on Image Processing*, vol. 19, no. 7, pp. 1808–1823, 2010.

[20] J. Goodman, *Introduction to Fourier Optics*. W. H. Freeman, 2017.

[21] A. V. Oppenheim, A. S. Willsky, S. H. Nawab, G. M. Hernandez' *et al.*, *Signals & systems*. Pearson Education, 1997.

[22] S. Pinilla, C. Noriega, and H. Arguello, "Stochastic truncated Wirtinger flow algorithm for phase retrieval using boolean coded apertures," in *2017 IEEE International Conference on Acoustics, Speech and Signal Processing (ICASSP)*. IEEE, 2017, pp. 6050–6054.

[23] D. Gross, F. Krahmer, and R. Kueng, "Improved recovery guarantees for phase retrieval from coded diffraction patterns," *Applied and Computational Harmonic Analysis*, vol. 42, no. 1, pp. 37–64, 2017.

[24] Dwivedi, A.K., Ghosh, S. & Londhe, N.D. Review and Analysis of Evolutionary Optimization-Based Techniques for FIR Filter Design. Circuits Syst Signal Process 37, 4409–4430 (2018).

[25] T. N. Davidson, "Enriching the Art of FIR Filter Design via Convex Optimization," in IEEE Signal Processing Magazine, vol. 27, no. 3, pp. 89-101, May 2010.

[26] X. Pan, C. Liu, and J. Zhu, "Coherent amplitude modulation imaging based on partially saturated diffraction pattern," *Optics Express*, vol. 26, no. 17, p. 21929, aug 2018.

[27] Y. Wu, Y. Zhou, G. Saveriades, S. Agaian, J. P. Noonan, and P. Natarajan, "Local shannon entropy measure with statistical tests for image randomness," *Information Sciences*, vol. 222, pp. 323–342, 2013, including Special Section on New Trends in Ambient Intelligence and Bio-inspired Systems.

[28] S. Behnia, A. Akhshani, H. Mahmodi, and A. Akhavan, "A novel algorithm for image encryption based on mixture of chaotic maps," *Chaos, Solitons & Fractals*, vol. 35, no. 2, pp. 408–419, 2008.

[29] C. E. Shannon, "A mathematical theory of communication," *The Bell System Technical Journal*, vol. 27, no. 3, pp. 379–423, 1948.

[30] H. Chang, Y. Lou, Y. Duan, and S. Marchesini, "Total variation–based phase retrieval for poisson noise removal," *SIAM Journal on Imaging Sciences*, vol. 11, no. 1, pp. 24–55, 2018.

[31] S. P. Boyd, N. Parikh, E. Chu, B. Peleato, and J. Eckstein, "Distributed optimization and statistical learning via the alternating direction method of multipliers," *Found. Trends Mach. Learn.*, vol. 3, pp. 1–122, 2011.

[32] R. Li and L. Cao, "Progress in phase calibration for liquid crystal spatial light modulators," *Applied Sciences*, vol. 9, no. 10, 2019.

[33] A. Bergeron, J. Gauvin, F. Gagnon, D. Gingras, H. H. Arsenault, and M. Doucet, "Phase calibration and applications of a liquid-crystal spatial light modulator," *Appl. Opt.*, vol. 34, no. 23, pp. 5133–5139, Aug 1995.